\documentclass[prd,twocolumn,nofootinbib]{revtex4}

\begin{document}

\title{TWO--TWISTOR DESCRIPTION OF MEMBRANE}

\author{Sergey  Fedoruk,${}^1$ Jerzy Lukierski${}^2$\footnote{Supported by  KBN
grant 1 P03B 01828}}

\affiliation{\vspace{0.5cm} ${}^1$Bogoliubov Laboratory of  Theoretical Physics, JINR,  141980 Dubna, Moscow Region, Russia \\
${}^2$Institute for Theoretical Physics, University of Wroc{\l}aw, pl. Maxa Borna 9, 50-204
Wroc{\l}aw, Poland}

\begin{abstract}
\vspace{0.5cm} \noindent We describe $D=4$ twistorial membrane in
terms of two twistorial three--dimensional world--volume fields. We
start with the $D$--dimensional $p$--brane generalizations of two
phase space string formulations:  the one with $p+1$ vectorial
fourmomenta, and the second with tensorial momenta of $(p+1)$-th
rank. Further we consider tensionful membrane case in $D=4$. By
using the membrane generalization of Cartan--Penrose formula we
express the fourmomenta by spinorial fields and obtain the
intermediate spinor--space-time formulation. Further by expressing
the world--volume dreibein and the membrane space-time coordinate
fields in terms of two twistor fields one obtains the purely
twistorial formulation. It appears that the action is generated by a
geometric three--form on two--twistor space. Finally we comment on
higher--dimensional ($D>4$) twistorial $p$--brane models and their
superextensions.

\end{abstract}
\vspace{2cm}

\maketitle

\section{Introduction}

Since long time the idea of twistor space (see e.~g.~\cite{PenMac})
as describing the basic geometric arena for the physical phenomena
is tested in various ways. In particular it is known that massless
relativistic particles can be equivalently described by free
one--twistor particle model (see e.~g.~\cite{Fer,Shir,BC,STV});
further it can be shown that massive relativistic particles with
spin require in twistorial approach the two--twistor
space~\cite{Hug,Bet,FedZim,BetAz,Az,FFLM}\footnote{One should add
that unconventional application of one--twistor geometry to massive
particles has been proposed in~\cite{Bars}.}. Further recently there
was derived the twistor space action for general Yang--Mills
(Euclidean) gauge theory~\cite{Mas1,Mas2} providing a breakthrough
in extending the twistor construction to non--self--dual field
theories. Important contribution to the twistor programme is also
provided by the proof that large class of perturbative amplitudes in
$N=4$ $D=4$ supersymmetric YM theory and conformal supergravity can
be derived by using tensionless superstring moving in supertwistor
space. In such approach (see e.~g.~\cite{Wit,BerWit}) the twistorial
classical string is described by two--dimensional $CP(3|4)$
$\sigma$--model with (twisted) $N=2$ world sheet SUSY. The
correspondence with the space--time picture was derived~\cite{Wit}
on the level of quantized supertwistor string, providing twistor
superstring field theory.
 By assuming the topological CS action for
the (Euclidean) twistor superstring field one can reduce all the
superstring excitations to the massless spectrum describing SUSY YM
and supergravity theories. On the other hand
 recently
in $D=4$ the close link between the tensionful string model (Nambu-Goto action) and twistor
geometry has been proposed on the classical level, with target space--time string
coordinates composite in terms of twistor string fields~\cite{FL,Tay}. It appears that in
such a model the twistorial target space in bosonic $D=4$ string model consists of a pair
of twistor string coordinates.

Recently in $M$--theory the (super)strings have lost their privileged role as the
candidates for the Theory of Everything, and higher--dimensional $p$--branes ($p>1$), in
particular membranes, were considered. In this paper we shall show that analogously to the
result presented in~\cite{FL} one can introduce in two--twistor target space the purely
twistorial membrane action. In the intermediate spinor--space-time formulation with
fundamental Lorentz spinors and space--time coordinates we shall get the membrane action
which can be also linked with the $p$--brane description in the framework of spinorial
harmonics~(\cite{BZ-p}; see also~\cite{BPSTV,BSV}). It should be added that the presence of
cosmological term in Polyakov type action for membrane permits to obtain the purely
twistorial model without the use of gauge fixing procedure, which was a necessary step in
string case~\cite{FL}.

For simplicity we shall describe our composite membrane models in detail in $D=4$, and
without supersymmetric extension. Taking however into consideration that the
multidimensional and supersymmetric extensions of twistors applied to twistorial string
formulations has been already considered (see for example~\cite{Wit,BerWit,BAM,Uv}), we
believe that this paper can be also useful e.~g. for the consideration of the composite
``$M$--theoretic'' $D=11$ supermembrane.

In order to achieve our goal firstly in Sect.~2 we shall show how to extend from string to
$p$--brane the two known phase space formulations of bosonic string theory: { the one
using vectorial momenta fields~\cite{Sieg2,Pav-PL} and the second using tensorial momenta
fields (see e.~g.~\cite{GG,AurS,GZ,Nieto})}. These two formulations are based on the
following two Liouville $(p+1)$--forms which define the $p$--brane momenta fields:

\begin{description}
  \item[a)] {Vectorial momenta} model:
\begin{equation}\label{Th-p-int}
\Theta^{(1)} = p_\mu \,d X^\mu \quad \rightarrow \quad \Theta^{(p+1)} = P_\mu \wedge d X^\mu
\end{equation}
where $P_\mu$ is the following $p$--form
\begin{equation}\label{P-form-int}
P_\mu =P_\mu^{\,m}\epsilon_{mn_1\ldots n_p} d \xi^{n_1} \ldots d \xi^{n_p} \,.
\end{equation}
It is easy to see that if $p=1$ we obtain the Siegel formula~\cite{Sieg2} for the string momenta one form
\begin{equation}\label{P-st-int}
P_\mu =P_\mu^{\,m} d \xi^{n} \epsilon_{mn} \,.
\end{equation}

  \item[b)] Tensorial momenta model:
\begin{equation}\label{tTh-p-int}
\Theta^{(1)}\! = p_\mu \,d X^\mu \, \rightarrow \, {\tilde\Theta}{}^{(p+1)}\! = P_{\mu_1
\ldots \,\mu_{p+1}} d X^{\mu_1} \wedge \ldots \wedge d X^{\mu_{p+1}}\,.
\end{equation}
If $p=1$ one obtains the tensorial string momenta $P_{\mu\nu}=-P_{\nu\mu}$
(\cite{GG,GZ}; for arbitrary $p$ see~\cite{Nieto}).
\end{description}

In Sect.~2 we shall consider $p$--branes with arbitrary $p$ and in
arbitrary space--time dimension $D$. We shall show that both phase
space formulations using the momenta fields~(\ref{P-form-int})
or~(\ref{tTh-p-int}) are equivalent to the $p$--brane
Dirac--Nambu--Goto action~\cite{Dirac,Nambu} as well as to the
$\sigma$--model action ($p$--brane extension {of
Howe--Tucker--Polyakov action for strings and
membranes~\cite{HT,Pol,Sug})}. Further, in Sect.~3 we consider for
the case $p=2$ and $D=4$ (fourdimensional membrane) the intermediate
spinor--space-time models for both phase space formulations. It
appears that in such a model the spinors are
constrained.\footnote{We propose alternative way of generating
constraints in comparison with the framework of Lorentz harmonic
approach~\cite{BZ-p,BPSTV,BSV}.} In Sect.~4 we shall consider purely
twistorial action for $D=4$ membrane. We show that in this action
the Lagrangian density is described by a canonical twistorial
three--form. In Sect.~5 we present an outlook: we comment on the
description of twistorial membranes in higher dimensions, their
supersymmetric extensions and present the remarks about the purely
twistorial $p$--branes ($p>2$).

\section{Two phase space formulations of the tensionful $p$--brane
in $D$--dimensional space--time}

The tensionful $p$--brane propagating in flat Minkowski space is described by the nonlinear
Dirac--Nambu--Goto action~\footnote{The indices $m,n=0,1,\ldots,p$ are vector world-sheet
indices; $\mu,\nu=0,1,\ldots, D-1$ is vector space--time ones. We use the following flat
metrics: $\eta^{ab}=(-,+,\ldots,+)$, $\eta^{\mu\nu}=(-,+,\ldots,+)$.}
\begin{equation}\label{action-NG}
S=-T\int d\,^{p+1}\xi\, \sqrt{- g}
\end{equation}
where $\xi^m=(\tau,\sigma^1,\ldots,\sigma^p)$ are the world--volume coordinates,
\begin{equation}\label{g-det}
g\equiv \det (g_{mn})
\end{equation}
and
\begin{equation}\label{ind-metr}
g_{mn}= \partial_m X^\mu\partial_n X_\mu
\end{equation}
is the induced metric on the ($p+1$)--dimensional $p$--brane volume, $T$ is the $p$--brane
tension. In the case $p=1$ the action~(\ref{action-NG}) is the Nambu--Goto action for
string~\cite{Nambu} whereas if $p=2$ the action~(\ref{action-NG}) is the Dirac action for
relativistic membrane~\cite{Dirac}.

Because the twistor coordinates replace the standard phase space variables, the transition
to twistorial formulation should be imposed on the Hamiltonian--like formulations. In the
case of tensionful $p$--branes~(\ref{action-NG}) there are known two Hamiltonian
descriptions.

\subsection{Phase space formulation with vectorial momenta}

Let us use firstly the vectorial momenta defined by the formula~(\ref{P-form-int}). The
corresponding action of the tensionful $p$--brane looks as follows~\cite{Pav-PL}
\begin{eqnarray}
S= \int  d^{p+1}\xi & \!\!\Big[ & \!\! P_\mu^{\,m}\partial_m X^\mu + {\textstyle
\frac{1}{2T}}(-h)^{-1/2}h_{mn}P_\mu^{\,m} P^{\,\mu\,n} + \nonumber\\
&& + {\textstyle \frac{T}{2}} (p-1) (-h)^{1/2} \Big]\,.  \label{act-s}
\end{eqnarray}
We note that last `cosmological' term in the action is absent if $p=1$
(string case).

Let us write down the equation of motion obtained after varying the world--volume metric
$h_{mn}$. Using $\delta h=hh^{mn}\delta h_{mn}= -hh_{mn}\delta h^{mn}$ one gets
\begin{equation}\label{3}
P^{\,m}_\mu P^{\,\mu\,n}-{\textstyle \frac{1}{2}} h^{mn} \Big[ \, h_{kl}P^{\,k}_\mu P^{\,l \mu}
+ (p-1)h T^2 \Big]=0\,.
\end{equation}

Further expressing $P^{\,m}_\mu$ in the action~(\ref{act-s}) by its equation of motion
($h^{mk}h_{kn}=\delta^{m}_{n}$)
\begin{equation}\label{2}
P^{\,m}_\mu= -T(-h)^{1/2}h^{mn}\partial_n X_\mu
\end{equation}
one obtains the $\sigma$--model action for the $p$--brane { of the
Howe--Tucker--Polyakov type}~\footnote{The string actions as $d=2$
world sheet gravity interacting with string coordinate fields were
originally proposed in~\cite{BrDes}, where as well the description
of spinning string using interacting $d=2$ supergravity is
presented.}
\begin{equation}\label{action-2}
S=-{\textstyle \frac{T}{2}}\int d^{p+1}\xi\, (-h)^{1/2} \Big[\, h^{mn}\partial_m X^\mu\partial_n X_\mu
-(p-1) \,\Big]
\end{equation}
where the variables $h^{mn}$ can be treated as independent ones. The equations of motions for $h^{mn}$ give
$$
\partial_m X^\mu\partial_n X_\mu - {\textstyle\frac{1}{2}}h_{mn} \Big[\, h^{kl}\partial_k X^\mu\partial_l X_\mu
-(p-1) \,\Big]=0
$$
and lead to (using $h^{mn}h_{mn}=p+1$)
\begin{equation}\label{h-g}
h_{mn}=  g_{mn}
\end{equation}
where $g_{mn}$ is the induced metric~(\ref{ind-metr}) on the $p$--brane
volume.\footnote{The relation~(\ref{h-g}) is unique for $p\neq 1$; for $p= 1$ one can
introduce in~(\ref{h-g}) an arbitrary local scaling factor.} After substitution
of~(\ref{h-g}) in~(\ref{action-2}) we obtain after simple algebraic calculation the
action~(\ref{action-NG}).

It can be shown that from the action~(\ref{act-s}) one can derive the $p+1$ Virasoro
constraints, generating the world--volume diffeomorphisms. We shall divide world--volume
indices $m,n = 0,1,\ldots ,p$ into one with zero value and remaining
$\underline{m},\underline{n} = 1,\ldots ,p$\,, that is $m =(0, \underline{m})$. The
equations~(\ref{2}) lead to expression of the `auxiliary space momenta'
\begin{equation}\label{sp-mom}
P^{\,\underline{m}}_\mu= - h_{0\underline{n}} \,
\underline{h}^{\underline{n}\,\underline{m}} \, P_\mu -T(-h)^{1/2}\,
\underline{h}^{\underline{m}\,\underline{n}} \,\partial_{\underline{n}} X_\mu
\end{equation}
where $P_\mu= P^{\,0}_\mu$
is true momentum and $(p\times p)$ matrix
\begin{equation}\label{h-und}
\underline{h}^{\underline{m}\,\underline{n}} = h^{\underline{m}\,\underline{n}} \,-\,
\frac{h^{0 \underline{m}}\,h^{0\underline{n}}}{h^{00}}
\end{equation}
is the inverse matrix for $(p\times p)$ space part of the world volume metric
($\underline{h}^{\underline{m}\,\underline{n}} \, h_{\underline{n}\,\underline{k}} =
\delta^{\underline{m}}_{\underline{k}}$). We note that
\begin{equation}\label{h-tozh}
\frac{1}{h^{00}} = h\, \det (\underline{h}^{\underline{m}\,\underline{n}})  \,, \qquad
\frac{h^{0 \underline{m}}}{h^{00}} =  - h_{0\,\underline{n}}
\,\underline{h}^{\underline{n}\,\underline{m}}  \,.
\end{equation}
Inserting~(\ref{sp-mom}) in the the action~(\ref{act-s}) we obtain the first order
Lagrangian
\begin{widetext}
\begin{equation}\label{L-ap1}
{\cal L} = P_\mu \dot X^\mu - {\textstyle\frac{\sqrt{-h}}{2T}} \, \det
(\underline{h}^{\underline{m}\,\underline{n}})  \,\Big( P_\mu P^\mu \Big) \,-\,
{\textstyle\frac{T\sqrt{-h}}{2}}\, \underline{h}^{\underline{m}\,\underline{n}} \, \Big(
\partial_{\underline{m}} X^\mu\partial_{\underline{n}} X_\mu \Big) \,- \,
h_{0\,\underline{n}} \,\underline{h}^{\underline{n}\,\underline{m}} \, \Big(
P_\mu\partial_{\underline{m}} X^\mu \Big) \,+\, {\textstyle\frac{T}{2}} \, (p-1)\sqrt{-h}
\,.  \nonumber
\end{equation}
Using the equality
$$
\underline{h}^{\underline{m}_1\,\underline{n}_1} = {\textstyle\frac{1}{(p-1)!}}\,\det
(\underline{h}^{\underline{m}\,\underline{n}})\,\epsilon^{\underline{m}_1 \underline{m}_2
\ldots \underline{m}_p}\,\epsilon^{\underline{n}_1 \underline{n}_2\ldots \underline{n}_p}\,
h_{\underline{m}_2\,\underline{n}_2} \ldots h_{\underline{m}_p\,\underline{n}_p}
$$
and the equations of motions~(\ref{h-g}) we obtain that third term in Lagrangian~(\ref{L-ap1})
takes the form
\begin{equation}\label{term3-ap}
-\, {\textstyle\frac{T\sqrt{-h}}{2}}\, \underline{h}^{\underline{m}\,\underline{n}} \,
\Big( \partial_{\underline{m}} X^\mu\partial_{\underline{n}} X_\mu \Big) = -\,
{\textstyle\frac{T\sqrt{-h}}{2}}\,\, p\, \det
(\underline{h}^{\underline{m}\,\underline{n}}) \, \det (g_{\underline{m}\,\underline{n}})
\,.
\end{equation}
If the equations~(\ref{h-g}) are valid we get
\begin{equation}\label{det-det}
\det (\underline{h}^{\underline{m}\,\underline{n}}) \,  \det (g_{\underline{m}\,\underline{n}}) =1
\end{equation}
and the last term in the Lagrangian~(\ref{L-ap1}) can be written as follows
\begin{equation}\label{term5-ap}
{\textstyle\frac{T}{2}} \, (p-1)\sqrt{-h} = {\textstyle\frac{T}{2}} \, (p-1)\sqrt{-h} \,
\det (\underline{h}^{\underline{m}\,\underline{n}}) \,  \det
(g_{\underline{m}\,\underline{n}})  \,.
\end{equation}
Inserting the expressions~(\ref{term3-ap}), (\ref{term5-ap}) in the
Lagrangian~(\ref{L-ap1}) we obtain the following standard form of the Lagrangian density in
first order formalizm
\begin{equation}\label{L-ap}
{\cal L} = P_\mu \dot X^\mu - {\textstyle\frac{\sqrt{-h}}{2}} \, \det
(\underline{h}^{\underline{m}\,\underline{n}})  \,\Big[ {\textstyle\frac{1}{T}}\, P_\mu
P^\mu  \,+\, T\, \det (g_{\underline{m}\,\underline{n}}) \Big] \,-  \, h_{0\,\underline{n}}
\,\underline{h}^{\underline{n}\,\underline{m}} \, \Big( P_\mu\partial_{\underline{m}} X^\mu
\Big) \,.
\end{equation}
\end{widetext}

We see that the formula~(\ref{L-ap}) describes the set of $p+1$ Virasoro constraints
\begin{eqnarray}
H_0 &\equiv& {\textstyle\frac{1}{T}}\, P_\mu P^\mu  \,+\, T\,
\det (g_{\underline{m}\,\underline{n}}) \approx 0 \,, \label{cons-0}\\
&& \nonumber \\
H_{\underline{m}} &\equiv& P_\mu\partial_{\underline{m}} X^\mu  \approx 0   \label{cons-n}
\end{eqnarray}
with the Lagrange multipliers which are some nonlinear functions of the world--volume metric $h_{{m}\,{n}}$.

Let us finally deduce from the relations~(\ref{3}) the $p$--brane mass--shell condition.
Multiplying~(\ref{3}) by $h_{mn}$ we get for $p>1$
\begin{equation}\label{mass-p}
h_{mn}P^{\,m}_\mu P^{\,\mu\,n} = -(p+1)\,h \,T^2
\end{equation}
or
\begin{equation}\label{mass-Lag}
{\textstyle\frac{1}{2T}}(-h)^{-1/2}h_{mn}P_\mu^{\,m} P^{\,\mu\,n} = {\textstyle\frac{1}{2}} (p+1) (-h)^{1/2} T\,.
\end{equation}
In string case ($p=1$) the contraction of l.~h.~s. of~(\ref{3}) with $h_{mn}$ is
identically vanishing and in any space--time dimension the string mass
condition~(\ref{mass-p}) is absent.

\subsection{Phase space formulation with tensorial momenta}

Other phase space formulation of the $p$--brane~(\ref{action-NG}) is the model with
tensorial momenta. It is obtained by the use of the Liouville
$(p+1)$--form~(\ref{tTh-p-int}). Such a formulation is directly related with the
interpretation of $p$--branes as describing the dynamical $(p+1)$--dimensional world volume
elements described by the following $(p+1)$--forms~\footnote{Total antisymmetric tensors
$\epsilon^{m_1 \ldots \, m_{p+1}}$, $\epsilon_{m_1 \ldots \, m_{p+1}}$ and $\epsilon^{a_1
\ldots \, a_{p+1}}$, $\epsilon_{a_1 \ldots \, a_{p+1}}$ have the components $\epsilon^{01
\ldots \, p}=1$, $\epsilon_{0 1 \ldots \, p}=-1$.}
\begin{widetext}
\begin{equation}\label{els}
d S^{\mu_1 \ldots \,\mu_{p+1}} = d X^{\mu_1} \wedge \ldots \wedge d
X^{\mu_{p+1}} =
\partial_{m_1} X^{\mu_1} \ldots \partial_{m_{p+1}} X^{\mu_{p+1}} \epsilon^{m_1 \ldots \, m_{p+1}} d^{p+1}\xi \,.
\end{equation}
The $p$--brane action with tensorial momenta looks as follows~\cite{Nieto}
\begin{equation}\label{tens-4}
S = {\textstyle \frac{2}{\sqrt{(p+1)!}}} \int  d^{p+1}\xi  \, \Bigg[ \, P_{\mu_1 \ldots
\,\mu_{p+1}}\, \Pi^{\mu_1 \ldots \,\mu_{p+1}}  - \Lambda\left( P^{\mu_1 \ldots \,\mu_{p+1}}
P_{\mu_1 \ldots \,\mu_{p+1}}  +  {\textstyle\frac{T^2}{4}}\right)  \Bigg] \end{equation}
\end{widetext}
where
\begin{equation}\label{Pi}
\Pi^{\mu_1 \ldots \,\mu_{p+1}} \equiv \epsilon^{m_1 \ldots \, m_{p+1}}
\partial_{m_1} X^{\mu_1} \ldots \partial_{m_{p+1}} X^{\mu_{p+1}}   \,.
\end{equation}
Expressing $P_{\mu_1 \ldots \,\mu_{p+1}}$ by its equation of motion, we get
\begin{equation}\label{p-mn}
P^{\mu_1 \ldots \,\mu_{p+1}}={\textstyle\frac{1}{2\Lambda}}\,\Pi^{\mu_1 \ldots \,\mu_{p+1}}\,.
\end{equation}
After substituting~(\ref{p-mn}) in the action~(\ref{tens-4}) we obtain
the $2(p+1)$-th order action
\begin{equation}\label{tens-5}
S={\textstyle\frac{1}{2\sqrt{(p+1)!}}}\int d^{p+1}\xi \left[ \, \Lambda^{-1} \Pi^{\mu_1
\ldots \,\mu_{p+1}} \Pi_{\mu_1 \ldots \,\mu_{p+1}} - \Lambda T^2 \, \right]\,.
\end{equation}
Eliminating the auxiliary field $\Lambda$ we obtain
\begin{equation}\label{action-NG-Pi}
S=-{\textstyle\frac{T}{\sqrt{(p+1)!}}}\int d\,^{p+1}\xi\, \sqrt{- \Pi^{\mu_1 \ldots
\,\mu_{p+1}} \Pi_{\mu_1 \ldots \,\mu_{p+1}}} \,.
\end{equation}
But the determinant of the matrix~(\ref{ind-metr}) is given by the formula
\begin{eqnarray}
\det (g_{mn}) &=& {\textstyle\frac{1}{(p+1)!}}\, \epsilon^{m_1 \ldots \,
m_{p+1}}\epsilon^{n_1 \ldots \, n_{p+1}} g_{m_1 n_1} \ldots g_{m_{p+1} n_{p+1}} \nonumber\\
&=& {\textstyle\frac{1}{(p+1)!}}\, \Pi^{\mu_1 \ldots \,\mu_{p+1}} \Pi_{\mu_1 \ldots
\,\mu_{p+1}} \,.  \label{Pi-det}
\end{eqnarray}
We see that the action~(\ref{action-NG-Pi}) is classically equivalent to the action~(\ref{action-NG}).

The formula~(\ref{tens-5}) is very useful if we wish to consider for any $p$ the
tensionless limit $T \rightarrow 0$. We obtain the formula~\cite{Schild,Lind,Nieto}
\begin{equation}\label{tensionless}
S_{T=0}={\textstyle\frac{1}{2\sqrt{(p+1)!}}}\int d^{p+1}\xi \,
{\textstyle\frac{1}{\Lambda}}\, \Pi^{\mu_1 \ldots \,\mu_{p+1}} \Pi_{\mu_1 \ldots
\,\mu_{p+1}}\,.
\end{equation}
The formula~(\ref{tensionless}) describes the $p$--brane counterpart of Brink--Schwarz
action for massless particle~\cite{BrSch}.

\section{Tensionful $D=4$ membrane ($p=2$) in intermediate spinor--space-time formulations}

\subsection{Formulation with vectorial momenta}

In order to obtain from the action~(\ref{act-s}) the intermediate spinor--space-time action
we should eliminate the fourmomenta $P_\mu^m$ by means of the the membrane generalization
of the Cartan--Penrose formula expressing fourmomenta as spinorial bilinears. On $d=3$
curved world volume it has the form~\footnote{$h_{mn}=e_m^a e_{n a}$ is a world--volume
metric, $e_m^a$ is the dreibein, $e_m^a e^m_b = \delta^a_b$, $e=\det(e_m^a)=\sqrt{-h}$. The
indices $a,b=0,1,2$, $m,n=0,1,2$ are $d=3$ vector indices; the indices $i,j=1,2$ are $d=3$
Dirac spinor indices.  We use bar for complex conjugate quantities,
$\bar{\lambda}_{\dot\alpha}^i =(\overline{\lambda_{\alpha i}})$, and tilde for
Dirac--conjugated $d=2$ spinors, $\tilde{\lambda}_{\dot\alpha}^i =
\bar{\lambda}_{\dot\alpha}^j(\rho^0)_j{}^i$.}
\begin{equation}\label{P-res}
P_{\alpha\dot\alpha}^{\,m}=e\,
\tilde{\lambda}_{\dot\alpha}\rho^m \lambda_{\alpha} =e
e^m_a\tilde{\lambda}_{\dot\alpha}^i(\rho^a)_i{}^j\lambda_{\alpha j}
\end{equation}
where $\lambda_{\alpha i}$ ($i=1,2$) are two $D=4$ commuting Weyl spinors, $(\rho^a)_i{}^j$
are $2\times 2$ Dirac matrices in three--dimensional Minkowski space--time ($\{\rho^a ,
\rho^b \} =2 \eta^{ab}$) and $\rho^m = e^m_a \rho^a$. After using~(\ref{P-res})  the second
term in the action~(\ref{act-s}) takes the form~\footnote{The matrix of charge conjugation
is the $2\times 2$ skew-symmetric tensor $\epsilon^{ij}$. For definiteness, we take
$\epsilon^{12}=1$. The rules of lifting and lowering the indices are following:
$a^i=\epsilon^{ij}a_j$, $a_i=\epsilon_{ij}a^j$ where
$\epsilon^{ij}\epsilon_{jk}=\delta^i_k$. Also, we use $D=4$ Penrose spinor--vector
conventions~\cite{PenMac} in which $P_\mu P^\mu =
P_{\alpha\dot\alpha}P^{\dot\alpha\alpha}$, $P_\mu X^\mu =
P_{\alpha\dot\alpha}X^{\dot\alpha\alpha}$ ($P_{\alpha\dot\alpha} = \frac{1}{\sqrt{2}}P_\mu
\sigma^\mu_{\alpha\dot\alpha}$ etc.)}
\begin{equation}\label{2-term}
{\textstyle \frac{1}{2T}}(-h)^{-1/2}h_{mn}P_\mu^{\,m} P^{\,n \mu}={\textstyle
\frac{3}{4T}}\,e\, (\lambda^{\alpha i}\lambda_{\alpha i})
(\tilde{\lambda}_{\dot\alpha}^j\tilde{\lambda}^{\dot\alpha}_j)
\end{equation}
where we used ${\rm Tr}(\rho^m \rho^n) =2h^{mn}$.

Let us recall the condition~(\ref{mass-Lag}) which is valid for $p>1$ i.~e. also for membrane.
In order to get consistency of~(\ref{mass-Lag}) and (\ref{2-term}) we should introduce the following
constraint on spinors $\lambda$
\begin{equation}\label{la-mass}
A \equiv (\lambda \lambda)(\tilde{\lambda}  \tilde{\lambda}) - 2\, T^2 =0
\end{equation}
(we use notations $(\lambda \lambda) \equiv(\lambda^{\alpha i}
\lambda_{\alpha i})$, $(\tilde{\lambda}  \tilde{\lambda})
\equiv(\tilde{\lambda}_{\dot\alpha}^i\tilde{\lambda}^{\dot\alpha}_i)$;
note that $\tilde{\lambda}_{\dot\alpha}^i\tilde{\lambda}^{\dot\alpha}_i
= \bar{\lambda}_{\dot\alpha}^i\bar{\lambda}^{\dot\alpha}_i$).
Putting~(\ref{P-res}) and
(\ref{2-term}) in~(\ref{act-s}) and imposing via Lagrange multiplier
the constraint~(\ref{la-mass}) we obtain the action
\begin{equation}\label{act-mix-mem}
S= \int  d^3\xi \, \Bigg[ e\left( \tilde{\lambda}_{\dot\alpha} \rho^m\!\lambda_{\alpha }\,
\partial_m X^{\dot\alpha\alpha}+ 2\, T \right) + \Lambda A \Bigg]
\end{equation}
which provides the intermediate spinor--space-time formulation of the membrane. Let us
observe that the action~(\ref{act-mix-mem}) is invariant under the following Abelian local
gauge transformation
\begin{equation}\label{phase-tr}
\lambda^\prime_{\alpha i} = e^{i\gamma}\lambda_{\alpha i}
\end{equation}
with real local parameter $\gamma(\xi)$. By fixing the gauge~(\ref{phase-tr}) we can
replace one real constraint~(\ref{la-mass})  by the following pair of
constraints~\footnote{Compare with the string case considered in~\cite{FL}.}
\begin{equation}\label{la-mass-2}
(\lambda \lambda)=(\tilde{\lambda}  \tilde{\lambda}) = \sqrt{2}\, T \,.
\end{equation}

\subsection{Formulation with tensorial momenta}

Let us consider the general action which has the following form
\begin{equation}\label{act-sim}
S= \int  d^{3}\xi \, \Big( e  e^m_a  Q_m^a +  2eT \Big)
\end{equation}
where $Q_m^a=Q_m^a(X,\lambda)$ do not depend on $e_m^a$. Using the relations
$$
e=-{\textstyle\frac{1}{3!}}\epsilon^{mnk}\epsilon_{abc} e_m^a e_n^b e_k^c\,, \qquad
ee^m_a=-{\textstyle\frac{1}{2}}\epsilon^{mnk}\epsilon_{abc}  e_n^b e_k^c
$$
we obtain the following equation of motion for $e^m_a$
\begin{equation}\label{eq-e}
e_m^a = -{\textstyle\frac{1}{T}}\, Q_m^a \,.
\end{equation}
Subsequently the action~(\ref{act-sim}) takes the following classically equivalent form
\begin{equation}\label{act-sim-fin}
S= - {\textstyle\frac{1}{6T^2}} \int  d^{3}\xi \, \epsilon_{abc} \epsilon^{mnk} Q_m^a Q_n^b Q_k^c\,.
\end{equation}
Choosing in the action~(\ref{act-sim})
\begin{equation}\label{Q-lX}
Q_m^a = (\tilde{\lambda}_{\dot\alpha}
\rho^a \lambda_{\alpha })\,
\partial_m X^{\dot\alpha\alpha}
\end{equation}
and after supplementing the constraint~(\ref{la-mass}) one gets our membrane
action~(\ref{act-mix-mem}). Inserting the formula~(\ref{eq-e}) we obtain~\footnote{The
Lagrange multiplier in~(\ref{act-mix-tens}) is obtained from the one in~(\ref{act-mix-mem})
by the rescaling $\Lambda \rightarrow \frac{2}{\sqrt{6}}\,\Lambda$.}
\begin{equation}\label{act-mix-tens}
S= {\textstyle\frac{2}{\sqrt{6}}} \int  d^{3}\xi \Big( P_{\alpha\dot\alpha, \beta\dot\beta,
\gamma\dot\gamma} \epsilon^{mnk} \partial_m X^{\dot\alpha\alpha} \partial_n
X^{\dot\beta\beta} \partial_k X^{\dot\gamma\gamma} + \Lambda A \Big)
\end{equation}
where tensorial momenta are composites in term of fundamental spinors~\footnote{The
coefficient in~(\ref{P-tens-l}) are chosen in consistency with the general $p$--brane
formula~(\ref{tens-4}).}
\begin{eqnarray}\label{P-tens-l}
P_{\mu\nu\lambda} &=& P_{\alpha\dot\alpha, \beta\dot\beta, \gamma\dot\gamma}  \\
&=& - {\textstyle\frac{1}{2\sqrt{6}\, T^2}} \,  \epsilon_{abc}
(\tilde{\lambda}_{\dot\alpha} \rho^a \lambda_{\alpha }) (\tilde{\lambda}_{\dot\beta} \rho^b
\lambda_{\beta }) (\tilde{\lambda}_{\dot\gamma} \rho^c \lambda_{\gamma }) \nonumber\,.
\end{eqnarray}

We see that the intermediate spinor--space-time action~(\ref{act-mix-tens}) with composite
tensorial momenta is obtained after the elimination of dreibein variables $e^a_m$. Taking
into account the relation
\begin{equation}\label{l-vec}
(\tilde{\lambda}_{\dot\alpha}\rho^a \lambda_{\alpha }) (\tilde{\lambda}^{\dot\alpha} \rho_b
\lambda^{\alpha }) = \delta^a_b\, T^2\,,
\end{equation}
following from~(\ref{la-mass}), and $\epsilon^{abc}\epsilon_{abc}=-3!$ we can show easily
that the tensor~(\ref{P-tens-l}) satisfies the membrane mass shell condition (compare with
the constraint in the action~(\ref{tens-4}))
\begin{equation}\label{P-sq}
P^{\mu\nu\lambda}P_{\mu\nu\lambda} = P^{\alpha\dot\alpha, \beta\dot\beta, \gamma\dot\gamma}
P_{\alpha\dot\alpha, \beta\dot\beta, \gamma\dot\gamma} = - {\textstyle\frac{T^2}{4}}  \,.
\end{equation}
We see that in the action~(\ref{act-mix-tens}) the condition~(\ref{P-sq}) follows from the
constraint~(\ref{la-mass}).

\section{Purely twistorial formulation of the membrane ($p=2$) in $D=4$ space--time}

Further we introduce second half of twistor coordinates $\mu^{\dot\alpha}_i$,
$\bar{\mu}^{\alpha i}$ by postulating the Penrose incidence relations generalized for $D=4$
membrane fields
\begin{equation}\label{Pen-inc}
\mu^{\dot\alpha}_i=X^{\dot\alpha\alpha}\lambda_{\alpha i} \,,\qquad \tilde{\mu}^{\alpha
i}=\tilde\lambda^i_{\dot\alpha} X^{\dot\alpha\alpha}\,.
\end{equation}
We shall rewrite the action~(\ref{act-mix-mem}) by taking into account the relations~(\ref{Pen-inc}).
Using the relations~(\ref{P-res}), (\ref{Pen-inc}) we obtain
\begin{eqnarray}\label{PX-lm}
P_{\alpha\dot\alpha}^{\,m}\partial_m X^{\dot\alpha\alpha} &=& e\,
\tilde{\lambda}_{\dot\alpha}\rho^m\!\lambda_{\alpha}\, \partial_m X^{\dot\alpha\alpha} \\
&=& {\textstyle \frac{1}{2}}\,e \, e_a^m \left(\tilde{\lambda}_{\dot\alpha}\rho^a\partial_m
\mu^{\dot\alpha} -\tilde{\mu}^\alpha\rho^a\partial_m\lambda_\alpha \right) + {\it c.c.}
\nonumber
\end{eqnarray}
If we introduce the four--component twistors ($A=1,\cdots,4$)
\begin{equation}\label{ZbZ}
Z_{Ai}=(\lambda_{\alpha i}, \mu_i^{\dot\alpha}), \qquad \tilde
Z^{Ai}=(\tilde\mu^{\alpha i}, -\tilde\lambda_{\dot\alpha}^{i} ),
\end{equation}
the relations~(\ref{PX-lm}) takes the form
\begin{eqnarray}\label{PX-ZbZ}
P_{\alpha\dot\alpha}^{\,m}\partial_m X^{\dot\alpha\alpha} &=& e\,
\tilde{\lambda}_{\dot\alpha}\rho^m\!\lambda_{\alpha}\, \partial_m X^{\dot\alpha\alpha} \\
&=& {\textstyle \frac{1}{2}}\, e\, e_a^m \left(\partial_m\tilde Z^{A}\rho^a Z_{A} - \tilde
Z^{A}\rho^a \partial_m Z_{A} \right)\,. \nonumber
\end{eqnarray}

Incidence relations~(\ref{Pen-inc}) with real space--time membrane
position field $X^{\dot\alpha\alpha}$ imply that the twistor field variables
satisfy the constraints
\begin{equation}\label{V}
V_i{}^j \equiv \lambda_{\alpha i}\tilde{\mu}^{\alpha j} - \mu^{\dot\alpha}_i
\tilde{\lambda}_{\dot\alpha}^j \approx 0
\end{equation}
which can be rewritten equivalently
\begin{equation}\label{V-Z}
V_i{}^j = Z_{Ai}\tilde Z^{Aj}\approx 0\,.
\end{equation}
We obtain the following membrane action~(\ref{act-mix-mem}) in twistor formulation with dreibein
\begin{eqnarray}
S= \int  d^{3}\xi \!&\! \Bigg[ \!&\! {\textstyle \frac{1}{2}}\, e\, e_a^m
\left(\partial_m\tilde
Z^{A}\rho^a Z_{A} - \tilde Z^{A}\rho^a \partial_m Z_{A} \right) + \nonumber\\
&& + \, 2e\, T  + \Lambda A +\Lambda_j{}^i V_i{}^j \Bigg]  \label{action-tw-mem}
\end{eqnarray}
where $\Lambda$ and $\Lambda_i{}^j$ are the Lagrange multipliers.
If we define the asymptotic twistors~\cite{PenMac,Hug}
\begin{equation}\label{I}
I^{AB}=\left(
\begin{array}{cc}
  \epsilon^{\alpha\beta} & 0 \\
  0 & 0
\end{array}
\right)\,, \qquad I_{AB}=\left(
\begin{array}{cc}
  0 & 0 \\
  0 & \epsilon^{\dot\alpha\dot\beta}
\end{array}
\right)
\end{equation}
one can introduce the following notation
\begin{eqnarray}
(\lambda\lambda) \equiv &  \lambda^{\alpha i} \lambda_{\alpha i} = \epsilon^{ij} I^{AB}
Z_{Ai}Z_{Bj} &\equiv (ZZ)\,, \label{ll-ZZ1}\\
(\tilde\lambda \tilde\lambda) \equiv & \tilde\lambda_{\dot\alpha}^{i}
\tilde\lambda^{\dot\alpha}_i = \epsilon_{ij} I_{AB} \tilde Z^{Ai}\tilde Z^{Bj} &\equiv
(\tilde Z\tilde Z) \label{ll-ZZ2}
\end{eqnarray}
and write down the fourlinear constraint~(\ref{la-mass}) in the following twistorial form:
\begin{equation}\label{Z-mass}
A \equiv (Z Z)(\tilde Z\tilde Z) - 2T^2 =0\,.
\end{equation}

We shall eliminate the dreibein $e^a_m$ by employing the formula~(\ref{eq-e}). The
action~(\ref{action-tw-mem}) correspond to the choice
\begin{equation}\label{Q-Z}
Q_m^a = {\textstyle \frac{1}{2}}\,
\left(\partial_m\tilde Z^{A}\rho^a
Z_{A} - \tilde Z^{A}\rho^a \partial_m Z_{A} \right)  \,.
\end{equation}
One gets the final action depending only on two twistorial fields $Z_{Ai}(\tau, \sigma^1,
\sigma^2)$ and suitably rescaled (in comparison with~(\ref{action-tw-mem})) the Lagrange
multipliers $\Lambda$, $\Lambda_j{}^i$:
\begin{widetext}
\begin{eqnarray}
S \!\!\!&=& \!\!\!-{\textstyle\frac{1}{48 T^2}}\, \int  d^{3}\xi \, \Bigg[ \epsilon_{abc}
\epsilon^{mnk} \left(\partial_m\tilde Z^{A}\rho^a Z_{A} \!-\! \tilde Z^{A}\rho^a \partial_m
Z_{A} \right)\! \left(\partial_n\tilde Z^{B}\rho^b Z_{B} \!-\! \tilde Z^{B}\rho^b
\partial_n Z_{B} \right)\! \left(\partial_k\tilde Z^{C}\rho^c
Z_{C} \!-\! \tilde Z^{C}\rho^c \partial_k Z_{C} \right) + \nonumber\\
&&  \qquad\qquad\qquad \qquad\qquad + \, \Lambda A  + \Lambda_j{}^i V_i{}^j  \Bigg] \,.
\label{act-pure-tw}
\end{eqnarray}
\end{widetext}

The model~(\ref{act-pure-tw}) describes the $D=4$ membrane in purely twistorial
formulation. Introducing three one--forms with world-volume--vectorial index
\begin{equation}\label{1-forms}
\Theta_{\!(1)}^{\,a} \equiv d\tilde Z^{A}\rho^a
Z_{A} \!-\! \tilde Z^{A}\rho^a d Z_{A}
\end{equation}
one can obtain the action~(\ref{act-pure-tw}) as induced on the membrane world volume by the following three--form
\begin{equation}\label{3-form}
\Theta_{\!(3)} = \epsilon_{abc}\, \Theta_{\!(1)}^{\,a} \wedge \Theta_{\!(1)}^{\,b} \wedge \Theta_{\!(1)}^{\,c}\,.
\end{equation}

\section{Outlook}

In  this paper we presented the new description of the twistorial membrane in $D=4$
space--time. We would like now to comment on two generalizations:
\begin{description}
  \item[i)] to $p$--branes with $p>2$ in arbitrary $D$--dimensional ($D>p+1$) space--time
  \item[ii)] to super--$p$--branes in higher dimensions.
\end{description}

The basic relation in the construction of twistorial formulation of the $p$--branes in
dimension $D$ is a suitable generalization of Cartan--Penrose formula~(\ref{P-res}). For
arbitrary $p$ and arbitrary $D$ such a formula looks as follows:
\begin{equation}\label{P-res-D}
P_{\mu}^{\,m}= e e^m_a\,\tilde{\lambda}^{\hat\alpha i}(\rho^a)_i{}^j\lambda_{\hat\beta j}\,
(\gamma_\mu)_{\hat\alpha}{}^{\hat\beta}
\end{equation}
where $\gamma_\mu$ are the Dirac matrices in $D$--dimensional space--time and $\rho^a$ are
Dirac matrices in $d=p+1$ dimensional space (tangent space to curved world volume geometry
of the $p$--brane). Thus the spinor $\lambda_{\hat\alpha i}$ has two indices: it is spinor
in $D$ dimensions with the components described by index $\hat\alpha$ and as well spinor in
$d$ dimensions with index $i$; we employ also the spinor $\tilde{\lambda}^{\hat\alpha i}$
which is a Dirac--conjugated spinor with respect to both indices:
$\tilde{\lambda}^{\hat\alpha i} =(\lambda_{\hat\beta j})^+
(\rho^0)^{ji}(\gamma^0)^{\hat\beta\hat\alpha}$. For general $D$ and $p$ we get that
$\hat\alpha=1,\ldots ,2^{[\frac{D}{2}]}$ and $i =1,\ldots ,2^{[\frac{p+1}{2}]}$. Thus in
order to obtain the composite $p$--brane momenta we must use at least $2^{[\frac{p+1}{2}]}$
twistors. For definite dimensions $p$ and $D$ we can further decrease the number of the
elementary spinor components $\lambda_{\hat\alpha i}$ by imposing consistently Majorana-,
Weyl- or Majorana--Weyl conditions.

Inserting~(\ref{P-res-D}) in the action~(\ref{act-s}) we shall obtain the intermediate
spinor--space-time formulation. Due to the mass--shell for the vectorial momenta
(see~(\ref{mass-p})) the elementary spinors $\lambda_{\hat\alpha i}$ will be constrained
(compare with~(\ref{la-mass}) and~(\ref{l-vec})). In order to get the purely twistorial
formulation of $p$--branes one has to introduce $D$--dimensional incidence
relation~(\ref{Pen-inc}) which provides the doubling of spinor components and lifts the
Lorentz spinors to twistors. It should be stress however that in general case the
$D$--dimensional incidence relation will introduce extended  $D$--dimensional space--time.
Only suitable use of the additional spinor structures (e.~g. quaternionic in $D=6$) and
imposition of the algebraic constraints in twistor space (e.~g. selecting only null
twistors lying on null hyperplanes) permits to obtain the incidence relations just with the
Minkowski space--time coordinates.

The extension of the twistorial formalism for bosonic $p$--branes to super--$p$--branes
requires the introduction of $p$--brane supertwistors. The techniques of
supersymmetrization of various twistorial $p$--brane models were already studied (see
e.~g.~\cite{BAM,Uv}). It should be also mentioned that our construction can be linked to
the analysis based on the use of Lorentz harmonics~\cite{BZ-p} as well as with the
formalism using $d=11$ BPS preons~\cite{BAIL,BAPV} described by generalized $OSp(1|64)$
supertwistor fields.

\begin{acknowledgments}
S.F. would like to thank Institute for Theoretical Physics, Wroc{\l}aw University  for kind
hospitality and a very friendly creative atmosphere. He would like to thank
Bogoliubov--Infeld program for financial support. The work of S.F. was partially supported
also by the RFBR grant 06-02-16684 and the grant INTAS-05-7928.
\end{acknowledgments}


\end{document}